\documentclass[journal]{new-aiaa}
\usepackage[utf8]{inputenc}

\usepackage{graphicx}
\usepackage{amsmath}
\usepackage[version=4]{mhchem}
\usepackage{siunitx}
\usepackage{longtable,tabularx}
\usepackage{hyperref}
\usepackage{siunitx}
\usepackage{graphicx}
\usepackage{tikz}
\usetikzlibrary{spy}
\usepackage{pgfplots}
\pgfplotsset{compat=newest}
\pgfplotsset{plot coordinates/math parser=false}
\usepackage{xcolor}
\usepackage{pdfpages}
\usetikzlibrary{plotmarks}
\usepackage{todonotes}
\usepackage{caption}
\usepackage{subfigure}

\title{Decoupling of Control and Force Objective in Adjoint-Based Fluid Dynamic Shape Optimization}

\author{Niklas K\"uhl\footnote{Research Associate, Institute for Fluid Dynamics and Ship Theory, niklas.kuehl@tuhh.de}  and Peter M. M\"uller\footnote{Student Assistant, Institute for Fluid Dynamics and Ship Theory, peter.mueller@tuhh.de}}
\affil{Hamburg University of Technology, Am Schwarzenberg-Campus 4, D-21075 Hamburg}
\author{Arthur St\"uck \footnote{Head of Department Simulation Frameworks,  Institute of Software Methods for Product Virtualization, Arthur.Stueck@dlr.de}}
\affil{German Aerospace Center (DLR), Institute of Software Methods for Product Virtualization, Zwickauer Strasse 46, D-01069 Dresden}
\author{Michael Hinze \footnote{Professor in Optimization of Complex Systems, Department of Mathematics, michael.hinze@uni-hamburg.de}}
\affil{University of Hamburg, Department of Mathematics, Bundesstrasse 55, D-20146 Hamburg}
\author{Thomas Rung \footnote{Professor for Fluid Dynamics, Institute for Fluid Dynamics and Ship Theory, thomas.rung@tuhh.de}}
\affil{Hamburg University of Technology, Am Schwarzenberg-Campus 4, D-21075 Hamburg}

\begin{document}
\newlength\figureheight
\newlength\figurewidth

\maketitle

\section*{Nomenclature}

\noindent(Nomenclature entries should have the units identified)

{\renewcommand\arraystretch{1.0}
\noindent\begin{longtable*}{@{}l @{\quad=\quad} l@{}}
$d_\mathrm{i}$ & optimization direction \\
$e_\mathrm{i}$ & general state constraint \\
$F_\mathrm{i}$ & general force \\
$G$ & gravitational constant \\
$G_\mathrm{i}$ & general gradient field \\
$g_\mathrm{i}$ & gravity field \\
$J$ & global objective functional \\
$j$ & local objective functional \\
$n_\mathrm{i}$ & normal vector \\
$p$ & fluid pressure \\
$\hat{p}$ & adjoint (fluid) pressure \\
$Q$ & residuum of mass balance \\
$\hat{Q}$ & residuum of adjoint mass balance \\
$R_\mathrm{i}$ & residuum of momentum balance \\
$\hat{R}_\mathrm{i}$ & residuum of adjoint momentum balance \\
$S_\mathrm{ij}$ &  strain rate tensor \\
$\hat{S}_\mathrm{ij}$ & adjoint strain rate tensor \\
$T$ & mapping of surfaces \\
$t$ & physical time \\
$u_\mathrm{i}$  & position of surface point \\
$V_\mathrm{i}$ & design velocity field \\
$v_\mathrm{i}$ & fluid velocity \\
$\hat{v}_\mathrm{i}$ & adjoint (fluid) velocity \\
$y_{\mathrm{i}}$ & general state variable  \\
$x_\mathrm{i}$ & general euclidean coordinate \\

$\Gamma$ & general domain boundary\\
$\delta_\mathrm{ij}$ & Kronecker delta \\
$\delta$ & general perturbation \\
$\epsilon$ & general increment \\
$\mu$ & fluid viscosity\\
$\rho$ & density\\
$\phi$ & general angle \\
\multicolumn{2}{@{}l}{Subscripts}\\
D & design boundary \\
i,j,k & general index for Einstein's summation convention\\
n & normal direction \\
O & objective boundary \\
$\Gamma$ & boundary based quantity \\
$\Omega$  & domain quantity \\
\multicolumn{2}{@{}l}{Superscripts}\\
l & local derivative\\
c & convective derivative \\
g & geometric derivative \\
$\tau$ & pseudo time\\
\end{longtable*}}

\section{Introduction}
This technical note discusses two alternatives for an adjoint-based derivative evaluation for fluid-dynamic forces on bodies within a node-based shape optimization problem \cite{kroger2015cad}. 
When attention is devoted to force objectives, the objective functional as well as the control are usually co-located, i.e. tangential and normal fluid-dynamic stresses are integrated along the body surface which serves as objective and design surface. This \emph{internal} objective functional evaluation introduces additional contributions to the total shape derivative that arise from the material derivative concept \cite{stuck2012adjoint, schmidt2013three}. An example for this additional contribution follows from a gravitational potential that introduces buoyancy forces  \cite{kroger2018adjoint}. These contributions can be deduced mathematically with the aid of the general shape calculus, since the integral bounds for determining the global force contain the controlled and thus variable shape \cite{jameson1995optimum, schmidt2010efficient, schmidt2013three}.
Alternatively, the fluid dynamic force on a body can be expressed by the global momentum loss  in an \emph{external} fashion \cite{squireyoung1937calculation}. This facilitates a decoupling between control and objective, where the latter is often evaluated at the boundaries of the computational domain. Mathematically, it is a matter of two different optimization problems (objective functional), that should be equivalent from a physical perspective.

The crucial point now is: The complexity of the objective-functional derivative expression is reduced by  decoupling control and objective surfaces, as geometric as well as convective contributions to the total derivative disappear due to the unperturbed control, and only the local contribution remains.  Different possibilities exist for the evaluation of the local contribution which can be subdivided into direct and indirect (adjoint) procedures \cite{giles2000introduction,peter2010numerical,stuck2012adjoint}, where the adjoint approach is deemed more efficient \cite{peter2010numerical, giles1997adjoint, giles2000introduction}. This note illustrates and discusses the external force evaluation from an adjoint perspective, which leads to a simplification of the final shape derivative.

Within the note, vectors and tensors are defined with reference to Cartesian coordinates and denoted by latin subscripts. Temporal information is indicated via superscripts and Einstein's summation  is used for latin subscripts.  

\section{Shape Calculus}
Assuming that an arbitrary shape $\Gamma$ is described by an infinite number of surface points $u_\mathrm{i}$. The transformation from an initial towards an alternative shape -- e.g. for optimization purposes -- can generally be performed in a \emph{pseudo time} $\tau$:
\begin{align}
\Gamma^{\tau} := \lbrace T^{\tau} ( u_\mathrm{i}^0 ): u_\mathrm{i}^0 \in \Gamma^0, \tau \geq  0 \rbrace \qquad \text{with} \qquad T^0 ( u_\mathrm{i}^0 ) = \Gamma^0
\end{align}
In comparison to the physical time $t$, the pseudo-time is a parent quantity. The mapping $T$ can be interpreted as a forward integration  following the gradient $G_\mathrm{i}^\tau$ in  pseudo time:
\begin{align}
T^{\tau} (  u_\mathrm{i}^0 ) := u_\mathrm{i}^0 + \tau G_\mathrm{i}^{\tau}  = u_\mathrm{i}^{\tau} \qquad \text{with} \qquad \tau \geq 0 \label{equ:pertu_of_ident},
\end{align}
frequently labelled as perturbation of identity. The temporal change of all surface coordinates can be described via the so called speed method:
\begin{align}
\frac{\mathrm{d} u_\mathrm{i}^{\tau}}{\mathrm{d} \tau} = V_\mathrm{i}^{\tau} \left( u_\mathrm{i}^{\tau} \right)  \qquad \text{with} \qquad \tau \geq 0 \label{equ:speed_metho} \, , 
\end{align}
where the velocity field $V_\mathrm{i}^{\tau}$ is often denoted as design velocity. Combining Eqns. (\ref{equ:pertu_of_ident}) and (\ref{equ:speed_metho}) yields $ \mathrm{d} T^{\tau} / \mathrm{d} \tau = G_\mathrm{i}^{\tau} = \mathrm{d} u_\mathrm{i}^{\tau} / \mathrm{d} \tau = V_\mathrm{i}^{\tau}$ for a steepest descent approach. We now aim to connect the design velocity with derivative information of a surface based objective function, e.g. a component of the force acting on a body:
\begin{align}
J = \int_{\Gamma_\mathrm{O}} j_\mathrm{\Gamma} \left( y_{\mathrm{j}} \right) \ \mathrm{d} \Gamma,
\end{align}
depending on the state variables $y_{\mathrm{j}}$ and the related boundary part $\Gamma_{\mathrm{O}}$. The shape derivative based on the introduced design velocity with state constraints is obtained by the concept of material derivative that decides between the initial and a perturbed domain and reads:
\begin{align}
\frac{\mathrm{d}}{\mathrm{d} \tau} J \left( u_\mathrm{i}^{\tau}, y_\mathrm{j} (u_\mathrm{i}^{\tau}) \right) \bigg|_{\tau = 0} 
&= \int_{\Gamma_\mathrm{O}} \underbrace{ \frac{\partial j_\Gamma}{\partial y_\mathrm{j}} y_\mathrm{j}^\prime  }_{\text{local: }\delta_\mathrm{u}^\mathrm{l} } \mathrm{d} \Gamma + \int_{\Gamma_D \cap \Gamma_O}  \bigg( \underbrace{ \frac{\partial j_\Gamma}{\partial x_\mathrm{j}} V_\mathrm{j}^0 }_{\text{conv.: }\delta_\mathrm{u}^\mathrm{c}}  + \underbrace{ j_\Gamma  \left( \frac{\partial V_\mathrm{j}^0}{\partial x_\mathrm{j}} - \frac{\partial V_\mathrm{j}^0}{\partial x_\mathrm{k}} n_\mathrm{j} n_\mathrm{k} \right)  }_{\text{geom.: }\delta_\mathrm{u}^\mathrm{g}} \bigg) \mathrm{d} \Gamma, \label{equ:total_shape_deriv} 
\end{align}
where the three parts are denoted as \emph{local}, \emph{convective} and \emph{geometric} derivative respectively \cite{schmidt2010efficient,stuck2012adjoint,schmidt2013three},
and $y_\mathrm{j}^\prime =\left( \mathrm{d} y_\mathrm{j} / \mathrm{d} u_\mathrm{i}^0 \right) V_\mathrm{i}^0 $ refers to the state derivative. 
A distinction between the habitat of the objective functional $\Gamma_\mathrm{O}$ and the control $\Gamma_\mathrm{D}$ is necessary: Both the convective as well as the geometric contribution need to be evaluated along the intersected area $\Gamma_\mathrm{D} \cap\Gamma_\mathrm{O} \neq \emptyset $. However, the most challenging part refers to the evaluation of the local contribution. For shape optimization problems the most efficient way to compute the local part is obtained from an adjoint analysis that circumvents a direct evaluation of the state derivative.

\section{Adjoint Evaluation of Local Derivatives}
The steady momentum and continuity equations for an incompressible, single-phase laminar Newtonian fluid -- that are solved for the velocity $v_\mathrm{i}$ and the pressure $p$  -- read:
\begin{alignat}{2}
R_i:&  \rho v_\mathrm{j} \frac{\partial v_\mathrm{i}}{\partial x_\mathrm{j}} + \frac{\partial }{\partial x_\mathrm{j}} \left[ p \delta_\mathrm{ij} - 2 \mu S_\mathrm{ij} \right] - \rho g_\mathrm{i} &&= 0 \label{equ:prima_momen} , \\
Q:& - \frac{\partial v_\mathrm{i}}{\partial x_\mathrm{i}} &&= 0 \label{equ:prima_mass} ,
\end{alignat}
where the unit coordinates, the strain rate tensor as well as the gravity vector are denoted by  $\delta_\mathrm{ij}$, $S_\mathrm{ij}$ and $g_\mathrm{i}$ respectively. 
Mind that the body force due to gravity can be scrambled into a generalized pressure via ${P} = p - \rho g_\mathrm{k} x_\mathrm{k}$.

The surface-based objective functional considered within this publication inheres local, geometrical and convective derivatives, whereby the local contribution can be evaluated efficiently by the adjoint approach. For this purpose, the objective functional is extended by the primal residuals to an augmented Lagrangian:
\begin{align}
\mathcal{L} = J +  \int_{\Omega} \left[ \hat{v_\mathrm{i}} R_\mathrm{i} + \hat{p} Q \right] \ \mathrm{d} \Omega  \label{equ:optim}
\end{align}
In order to derive the adjoint field equations, the partial derivative with respect to the local flow variations has to vanish in all directions ($\frac{\partial \mathcal{L}}{\partial v_\mathrm{i}} \frac{\mathrm{d} v_\mathrm{i}}{\mathrm{d} \tau} = \delta_{v_\mathrm{i}} \mathcal{L} \cdot \delta v_\mathrm{i} = 0 $, $ \frac{\partial \mathcal{L}}{\partial p} \frac{\mathrm{d} p}{\mathrm{d} \tau} =\delta_{p} \mathcal{L} \cdot \delta p = 0$), resulting in the following set of adjoint field equations:
\begin{alignat}{2}
\hat{R_\mathrm{i}}:&  -\rho v_\mathrm{j} \frac{\partial \hat{v}_\mathrm{i}}{\partial x_\mathrm{j}} +  \rho \hat{v}_\mathrm{j} \frac{\partial v_\mathrm{j}}{\partial x_\mathrm{i}}  + \frac{\partial}{\partial x_\mathrm{j}} \left[ \hat{p} \delta_\mathrm{ij} - 2 \mu \hat{S}_\mathrm{ij} \right] &&= 0 \label{equ:adjoi_momen} , \\
\hat{Q}:& -\frac{\partial \hat{v_\mathrm{i}}}{\partial x_\mathrm{i}} &&= 0 \, .  \label{equ:adjoi_mass}
\end{alignat}
Interestingly, compared to their primal counterpart the adjoint set of equations do not inhere any body-force term. The boundary conditions are derived from the boundary parts of the variation of the Langrangian:
\begin{align}
\delta_{v_\mathrm{i}} \mathcal{L} \cdot \delta v_\mathrm{i} &= \int_{\Gamma} \delta v_\mathrm{i} \left[ \rho \hat{v}_\mathrm{i} v_\mathrm{j} n_\mathrm{j} + 2 \mu \hat{S}_\mathrm{ij} - \hat{p} n_\mathrm{i} \right] - 2 \mu \left( \delta S_\mathrm{ij} \right) \hat{v}_\mathrm{i} n_\mathrm{j} \mathrm{d} \Gamma \\
 &+ \int_{\Gamma_\mathrm{O}} \delta v_\mathrm{i} \frac{\partial j_\Gamma  }{\partial v_\mathrm{i}} \mathrm{d} \Gamma \overset{!}{=} 0 \qquad  \forall \delta v_\mathrm{i}  \label{equ:local_veloc_varia} \\
\delta_{p} \mathcal{L} \cdot \delta p &= \int_{\Gamma} \delta p \ \hat{v}_\mathrm{i} \mathrm{d} \Gamma \\
&+ \int_{\Gamma_\mathrm{O}} \delta p  \frac{\partial j_\Gamma  }{\partial p}  \mathrm{d} \Gamma \overset{!}{=} 0 \qquad  \forall \delta p \label{equ:local_press_varia}
\end{align}
Note that $\Gamma_\mathrm{O}$ does not necessarily coincide with $\Gamma_\mathrm{D}$, which
 will be examined in more detail in the following section. With the aid of remaining optimality criteria, one can derive a local surface-based sensitivity rule along the design surface:
\begin{align}
\delta_\mathrm{u}^\mathrm{l} J  = -\int_{\Gamma_\mathrm{D}} \mu \frac{\partial v_\mathrm{i}}{\partial x_\mathrm{j}} \frac{\partial \hat{v_\mathrm{i}}}{\partial x_k} n_\mathrm{j} n_\mathrm{k} \mathrm{d} \Gamma \label{equ:final_sensi}.
\end{align}
Obviously, the local derivative is determined without directly differentiating the state into the direction of the control.

\section{Interior vs. Exterior Evaluation of Force Objectives}
\label{sec4}
\begin{figure}[h]
\subfigure[]{
\centering
\includegraphics[scale=1]{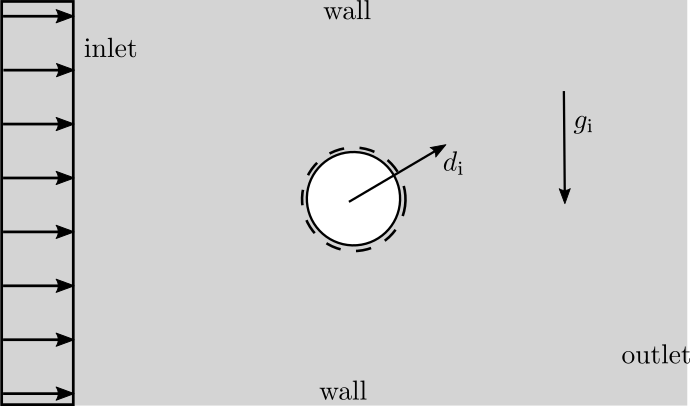}
}
\hspace{1cm}
\subfigure[]{
\centering
\includegraphics[scale=1]{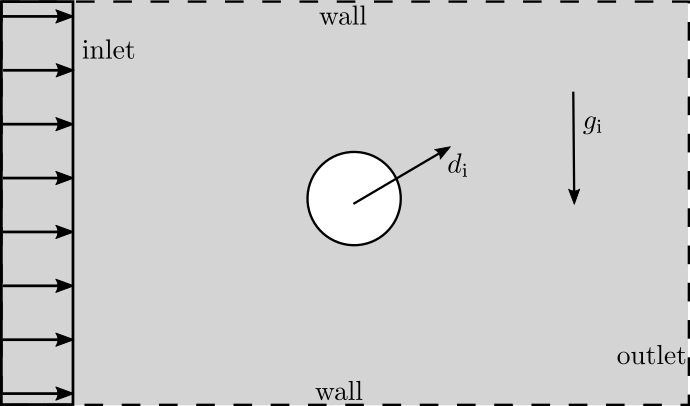}
}
\caption{Evaluation of forces acting on a cylinder: a) interior evaluation using surfaces stresses ($\Gamma_\mathrm{O} \equiv \Gamma_\mathrm{D}$) and b) exterior evaluation balancing the fluxes 
($\Gamma_\mathrm{D} \cap\Gamma_\mathrm{O} = \emptyset $). Integration areas are indicated by dashed lines that also hint on the habitat of the surface objective $j_\mathrm{\Gamma}$. 
}
\label{fig:inter_vs_exter}
\end{figure}

The section aims to convey that there is more than one answer in the sense of the adjoint approach to the same question. For this purpose a steady,  incompressible, laminar single-phase flow around a body is examined. The objective functional refers to the component of the flow-induced forces acting on the body in the direction $d_i$, i.e.  $J = d_\mathrm{i} F_\mathrm{i} $. The force can be determined in two different ways, referred to as interior (a) and exterior (b) strategy. Along all inlets and walls, the velocity is prescribed and zero normal-gradients are assumed for the velocity at the outlet.
Constant normal-gradients are used for the pressure to account for hydrostatics,
\begin{enumerate}
\item[(a)] Referring to  Fig. \ref{fig:inter_vs_exter}a, the projected force can be determined by integrating the stress components along the body surface and then projecting them into the direction of the objective force component:
\begin{align}
J =  \int_{\Gamma_\mathrm{O}} \left[  \left( p - \rho g_\mathrm{k} x_\mathrm{k} \right) \delta_{\mathrm{ij}} - 2 \mu S_{\mathrm{ij}} \right] n_\mathrm{j} d_\mathrm{i} \mathrm{d} \Gamma  . \label{equ:intern_objec}
\end{align}
In this classical example, the design surface coincides with the objective surface. This requires the evaluation of the complete derivative, i.e. all three terms of Eqn. (\ref{equ:total_shape_deriv}). Eliminating the local pressure variations in (\ref{equ:local_press_varia}) in combination with the force objective yields the following boundary conditions for the adjoint velocity:
\begin{align}
\hat{v}_\mathrm{i} = -d_\mathrm{i} \qquad \text{on cylinder } \left( \Gamma_\mathrm{O} = \Gamma_\mathrm{D} \right) \qquad \text{and} \qquad \hat{v}_\mathrm{i} = 0 \qquad \text{on ext. walls and inlet } 
\, . 
\end{align}
The complete set of adjoint boundary conditions is collected in Tab. \ref{tab:inter_vs_exter}. In line with the primal problem, the von-Neumann boundary conditions for the adjoint pressure are numerically motivated. 

\item[(b)] The force objective can also be examined by the momentum loss inside the computational domain (cf. Fig. \ref{fig:inter_vs_exter}b). The latter is evaluated by  integrating along the far field boundaries, viz.
\begin{align}
J =  \int_{\Gamma_\mathrm{O}} \left[  2 \mu S_{\mathrm{ij}} + \left( \rho g_\mathrm{k} x_\mathrm{k} - p \right) \delta_{\mathrm{ij}} - \rho v_\mathrm{i} v_\mathrm{j}  \right] n_\mathrm{j} d_\mathrm{i} \mathrm{d} \Gamma. \label{equ:exter_objec}
\end{align}
Since the objective functional is now decoupled from the design surface, only the local portion of the shape derivative needs to be considered. Eliminating the local pressure variations in (\ref{equ:local_press_varia}) in combination with the force objective  yields a different set of boundary conditions for the adjoint velocity 
 (see also Tab. \ref{tab:inter_vs_exter}):
\begin{align}
\hat{v}_\mathrm{i} = 0 \qquad \text{on cylinder } \left( \Gamma_\mathrm{D} \cap\Gamma_\mathrm{O} = \emptyset\right) \qquad \text{and} \qquad \hat{v}_\mathrm{i} = d_\mathrm{i} \qquad \text{on ext. walls and inlet } 
\end{align}
\end{enumerate}

\begin{table}[h]
\centering
\subtable[]{
\begin{tabular}{c||c|c||c|c}
boundary type & $v_\mathrm{i}$ & $p$ &  $\hat{v}_\mathrm{i}$  & $\hat{p}$ \\
\hline
\hline
inlet &$v_\mathrm{i} = \mathrm{const.}$ &$\frac{\partial p}{\partial \mathrm{n}} = \mathrm{const.}$& $\hat{v}_\mathrm{i} = 0$ & $\frac{\partial \hat{p}}{\partial \mathrm{n}} = 0$  \\
\hline
wall $\Gamma \backslash \Gamma_O$& $v_\mathrm{i} = 0$ &$\frac{\partial p}{\partial \mathrm{n}} = \mathrm{const.}$& $\hat{v}_\mathrm{i} = 0$ &  $\frac{\partial \hat{p}}{\partial n} = 0$ \\
\hline
wall $\Gamma \subset \Gamma_O$&$v_\mathrm{i} = 0$ &$\frac{\partial p}{\partial \mathrm{n}} = \mathrm{const.}$& $\hat{v}_\mathrm{i} = -d_\mathrm{i}$ &  $\frac{\partial \hat{p}}{\partial \mathrm{n}} = 0$ \\
\hline
outlet&$\frac{\partial v}{\partial \mathrm{n}} = 0$ &$\frac{\partial p}{\partial \mathrm{n}} = \mathrm{const.}$& $\hat{v}_\mathrm{i} = 0$ & $\hat{p} n_\mathrm{i}= 2 \mu \hat{S}_\mathrm{ij} n_\mathrm{j}$  \\
\end{tabular}
}
\hspace{1mm}
\subtable[]{
\begin{tabular}{ c|c}
 $\hat{v}_\mathrm{i}$  & $\hat{p}$ \\
\hline
\hline
  $\hat{v}_\mathrm{i} = d_\mathrm{i} $ & $\frac{\partial \hat{p}}{\partial \mathrm{n}} = 0$  \\
\hline
 $\hat{v}_\mathrm{i} = d_\mathrm{i} $ &  $\frac{\partial \hat{p}}{\partial n} = 0$ \\
\hline
  $\hat{v}_\mathrm{i} = 0$ &  $\frac{\partial \hat{p}}{\partial \mathrm{n}} = 0$ \\
\hline
 $\hat{v}_\mathrm{i} = d_\mathrm{i}$ & $\hat{p} n_\mathrm{i}= \rho d_\mathrm{i} v_\mathrm{j} n_\mathrm{j} + 2 \mu \hat{S}_\mathrm{ij} n_\mathrm{j}$  \\
\end{tabular}
}
\caption{Adjoint boundary conditions for the interior (a) and the exterior (b) evaluation of forces.}
\label{tab:inter_vs_exter}
\end{table}

The different boundary conditions reveal an additional observation: While in the interior case (a) a non-zero [zero] adjoint velocity sticks to the body to be optimized [far field], this is exactly the other way around in the exterior case (b).  The variable form of appearance resembles the transport theorem formulated in an Arbitrary Lagrangian Eulerian (ALE) frame of reference. Formulation (a) obviously corresponds to a moving grid approach, whereas  (b) is similar to an Eulerian description.

\section{Application}
\begin{figure}[h]
\centering
\includegraphics[scale=1]{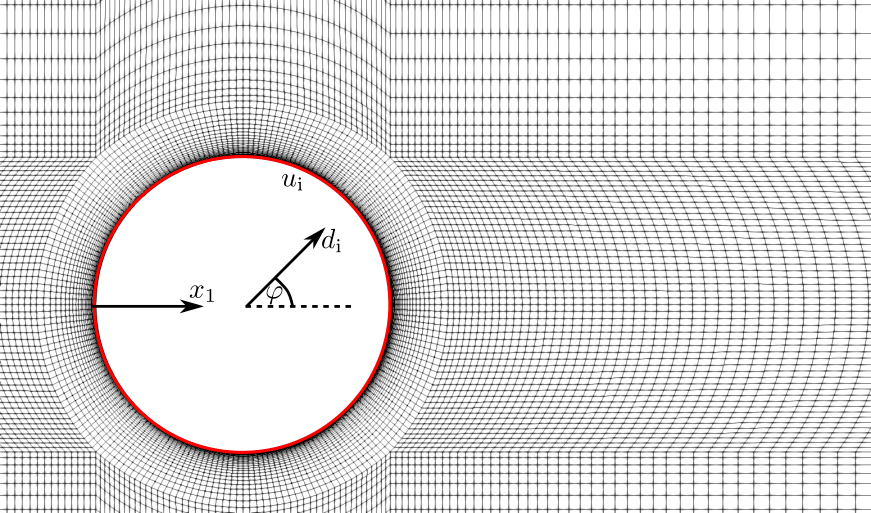}
\caption{Detail of the computational mesh used for the cylinder study at  $\mathrm{Fn} = 0.25$ and $\mathrm{Re} = 20$.}
\label{fig:comp_mesh}
\end{figure}

In this section the effect of decoupling control and objective functional is compared against the traditional coupled approach. For this purpose, we compute a simple but illustrative example, which refers to a steady, laminar, single-phase flow of an incompressible fluid around a cylinder under the influence of gravity. 
The Reynolds- and Froude-numbers are assigned to $\mathrm{Re} = 20$ and $\mathrm{Fn} = 0.25$ based on the constant inflow velocity and the diameter of the cylinder. 
Fig.  \ref{fig:comp_mesh} depicts the utilized mesh in the vicinity of the cylinder, which is discretized with 200 surface patches along the circumference. The boundary layer is fully resolved and the dimensionless near-wall spacing reaches down to $y^+ \approx O(10^{-2})$. The gravity vector $g_\mathrm{i}$ acts perpendicular to the approach flow direction.
A conventional, pressure-based, second-order accurate Finite-Volume scheme for a cell-centred variable arrangement is employed to approximate the partial differential equations of the primal and adjoint systems \cite{rung2009challenges}. 
Different objective force directions will be investigated for both evaluation strategies.  
Supplementary, Finite-Difference studies are performed for the exterior evaluation to verify the adjoint sensitivities. Note that a FD-study of the interior evaluation provides exactly the same result complemented by the related change of hydrostatic forces due to the variation of the fluid mass associated with the variation of the design surface. This is usually irrelevant in single-phase flows, fortunately not seen along the exterior boundaries and also not noticed by the adjoint single-phase system.   

\subsection{Comparison of Evaluations}
Evaluating the fluid forces for the interior approach  (\ref{equ:intern_objec}), the shape derivative includes all three contributions. The local change of the force objective functional is determined with the aid of the adjoint system using the boundary conditions illustrated in Table  \ref{tab:inter_vs_exter}. The convective contribution follows directly from the hydrostatic term $\rho g_\mathrm{k} x_\mathrm{k}$. Thanks to the structure theorem of Hadamard-Zol\'esio \cite{hadamard1968memoire, zolesio2011shapes}, only the normal component of the design velocity field influences the derivative ($ V_\mathrm{i}  = \frac{\mathrm{d} u_\mathrm{n}}{\mathrm{d} \tau} n_\mathrm{i}$). If curvature effects are neglected one arrives at:
\begin{align}
\delta_\mathrm{u}^c J = \int_{\Gamma_\mathrm{D} \cap \Gamma_\mathrm{O}}  \frac{\mathrm{d} u_\mathrm{n}}{\mathrm{d} \tau} \frac{\partial j_\Gamma}{\partial x_\mathrm{i}} n_\mathrm{i} \mathrm{d} \Gamma = \int_{\Gamma_\mathrm{D} \cap \Gamma_\mathrm{O}} \frac{\mathrm{d} u_\mathrm{n}}{\mathrm{d} \tau} \frac{\partial  ( \rho g_\mathrm{k} x_\mathrm{k} )}{\partial x_\mathrm{i}}  d_\mathrm{i} \mathrm{d} \Gamma = \rho g_\mathrm{k} d_\mathrm{k}  \int_{\Gamma_\mathrm{D} \cap \Gamma_\mathrm{O}} \frac{\mathrm{d} u_\mathrm{n}}{\mathrm{d} \tau} \mathrm{d} \Gamma 
\end{align}
The derivative inheres a convective contribution if and only if the force direction sees gravity (i.e., $g_\mathrm{i} d_\mathrm{i} \neq 0$). Interestingly, if the normal design velocity is approximated via a perturbation constant ($\frac{\mathrm{d} u_\mathrm{n}}{\mathrm{d} \tau} \approx \epsilon$), the latter cancels out as demonstrated by the local Finite-Difference (FD) results, viz.  
\begin{align}
\delta_{u = \mathrm{\tilde{u}}}^\mathrm{c} J = \frac{J \left( \tilde{u}_\mathrm{i} + \epsilon n_\mathrm{i} \right) - J \left( \tilde{u}_\mathrm{i} - \epsilon n_\mathrm{i} \right)}{2 \epsilon} = \frac{1 }{2 \epsilon} \left[  \int_{\Gamma_\mathrm{D}^\tau \cap \Gamma_\mathrm{O}} \rho g_\mathrm{k} d_\mathrm{k} \epsilon \mathrm{d} \Gamma - \int_{\Gamma_\mathrm{D}^\tau \cap \Gamma_\mathrm{O}} \rho g_\mathrm{k} d_\mathrm{k} \left( - \epsilon \right) \mathrm{d} \Gamma \right] = \rho g_\mathrm{k} d_\mathrm{k}.
\label{extraterm}
\end{align}

The geometric contribution accounts for the tangential change of the surface normals. If the boundary $\Gamma_\mathrm{O} \cap \Gamma_\mathrm{D}$ can be described locally as the graph of a $C^{2}$ function and considering the tangential Stokes formula, the tangential change agrees with the mean curvature. In case of a circular cylinder, this leads to a constant and the complete derivative of the objective inheres a Froude, curvature and Reynolds term:   
\begin{align}
\delta_\mathrm{u} J = \left( \delta_\mathrm{u}^\mathrm{c} + \delta_\mathrm{u}^\mathrm{g} + \delta_\mathrm{u}^\mathrm{l}  \right) J &= \int_{\Gamma_\mathrm{O} \cap \Gamma_\mathrm{D}} \left( \rho g_\mathrm{k} d_\mathrm{k} + C \right)  \mathrm{d} \Gamma -  \int_{\Gamma_\mathrm{D}}  \mu \frac{\partial v_\mathrm{i}}{\partial x_\mathrm{j}} \frac{\partial \hat{v_\mathrm{i}}}{\partial x_k} n_\mathrm{j} n_\mathrm{k} \mathrm{d} \Gamma  \qquad C \in \mathbb{R} 
\end{align}
Assessing the same problem on the same mesh for the exterior evaluation strategy,
yields to the same adjoint field equations with alternative boundary conditions (cf. Tab. \ref{tab:inter_vs_exter}). Based on the general shape calculus, it is expected that convective and geometric derivatives disappear and only the local derivative contributes. 

\subsection{Numerical Results}
\begin{figure}[h]
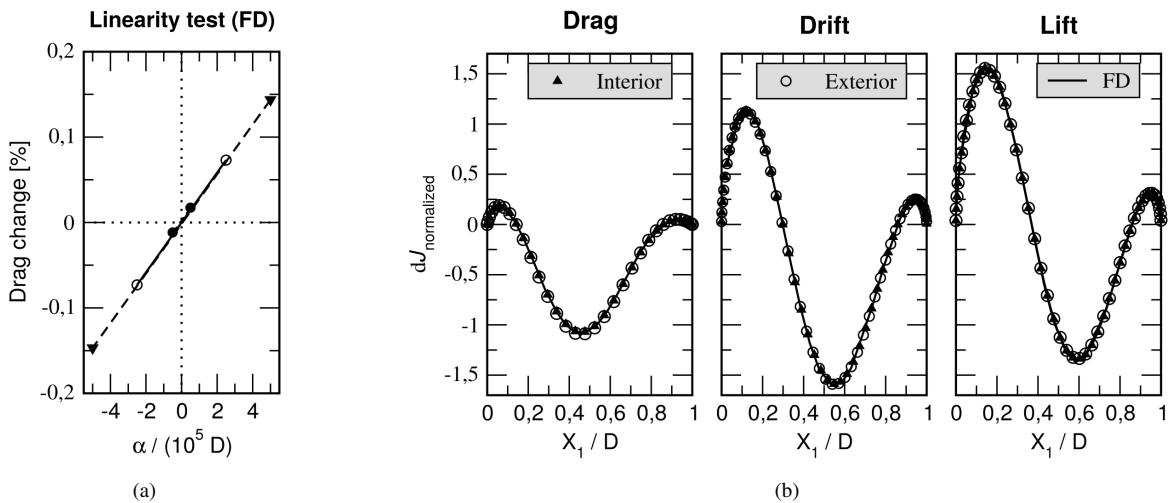

\subfigure[]{
\centering
\includegraphics[scale=0.6]{Picture2_a.pdf}
}
\hspace{1cm}
\subfigure[]{
\centering
\includegraphics[scale=0.6]{Picture2_b.pdf}
}
\caption{
Assessment of exterior force objective functional results: a) Comparison of three exemplary drag-objective responses to three central FD perturbations of a single surface patch based on consecutively doubled  perturbation sizes. b) Normalised local shape derivative assessed for three different objective directions, i.e. $\varphi = 0^\circ$ (drag), $\varphi = 45^\circ$ (drift) and $\varphi = 90^\circ$ (lift) compared with corresponding FD results.}
\label{fig:exter_deriv_90}
\end{figure}

To reliably compare results from Finite-Differences against adjoint sensitivities, it is crucial that the perturbation sizes lie within a regime dominated by a linear response. Fig. \ref{fig:exter_deriv_90}(a) shows the exemplary answers of the drag functional ($\varphi = 0^\circ$, cf. Fig. \ref{fig:comp_mesh}) to an initial and two consecutively doubled perturbations at one particular surface patch. All three results pass almost exactly through the origin, which indicates no significant non-linear effects. Obviously, the chosen perturbations comply fairly well with the linearity constraint, which was also verified for all other FD results of this note.
The local derivative for the investigated force directions (drag, drift, lift) is displayed in Fig. \ref{fig:exter_deriv_90}(b), which indicates that the FD-results and both adjoint sensitivities are in fair agreement for all three scenarios. This observation underscores potential merits of decoupling design and objective regimes by virtue of vanishing convective derivatives. 

\section{Conclusion}
We discuss exterior and classical interior alternatives for evaluating fluid flow induced forces on bodies. The discussion aims at a reduction of the total shape derivative, achieved through a decoupling of control and objective in the exterior approach. In this case, geometric as well as convective contributions to the shape derivative vanish. Convective contributions depend on primal physics and may disappear, which is not the case for geometric components. The latter can be interpreted as curvatures immanent to industrial applications. The remaining local derivative of the objective functional can be determined efficiently with an adjoint system, that differs to the classical approach in its boundary conditions only and resembles an ALE strategy. A two-dimensional flow exposed to gravity illustrates the features of the exterior approach, whereby carefully derived derivatives from a second order Finite-Difference study were used to validate the results.

\section*{Acknowledgements}
The current work is a part of the research project "Drag Optimisation of Ship Shapes" funded by the German Research Foundation (DFG, Grant No. RU 1575/3-1). This support is gratefully acknowledged by the authors.

\bibliography{./library.bib}

\end{document}